\documentclass[aps,pra,preprint,twocolumn,10pt]{revtex4}
\usepackage{graphicx}
\usepackage{revsymb}
\usepackage{amsthm, amssymb, amsmath}
\usepackage[matrix,arrow]{xy}

\newtheoremstyle{theorem}{5mm}{3mm}{\itshape}{}{\bfseries}{.}{1em}
		  {\thmname{#1}\thmnumber{ #2}\thmnote{ (#3)}}

\newtheoremstyle{dfn}{5mm}{3mm}{}{}{\bfseries}{.}{1em}{}

\newtheoremstyle{proof}{0mm}{3mm}{}{}{\bfseries\itshape}{.}{1em}
		{\thmname{#1}\thmnumber{ #2}\thmnote{ #3}}
  
\theoremstyle{theorem}
\newtheorem{theorem}{Theorem}
\newtheorem{lemma}[theorem]{Lemma}

\theoremstyle{dfn}
\newtheorem{definition}[theorem]{Definition}

\theoremstyle{proof}
\newtheorem*{pf}{Proof}
\renewenvironment{proof}{\begin{pf}}{\hfill\qed\end{pf}}

\def\idop{\openone}

\def \ge		{\geqslant}      
\def \le		{\leqslant}      
\renewcommand\preceq {\preccurlyeq}
\renewcommand\succeq {\succcurlyeq}
\def \N			{\mathbb N}      
\def \e			{\mathrm e}      
\def \dom		{\textrm{\textup{dom}}}
\def \img      {\text{\textrm{\textup{img}}}\:}	
\def \implies			{\Longrightarrow}					
\def \iff				{\Longleftrightarrow}				
\def \ox				{\otimes}              				

\def \union				{\,\cup\,}       
\def \inter				{\,\cap\,}       

\def \vide		{\varnothing}

\def \to		{\longrightarrow}                                 
\def \smallto		{\rightarrow}					  

\def \setminus		{\smallsetminus}				  
\def \subset		{\subseteq}					  
\DeclareMathSymbol	{\prsubset}{\mathrel}{symbols}{26}		  
\DeclareMathSymbol	{\prsupset}{\mathrel}{symbols}{27}		  
\DeclareMathSymbol	{\varphi}{\mathalpha}{letters}{39}		  

\newcommand		\ens[1]	{\left\{ #1 \right\}}		       
\newcommand		\abs[1]	{\left| #1 \right|}		       

\newcommand\ket[1]		{\left\lvert#1\right\rangle\mspace{-1.5mu}}						
\newcommand  \paren[1]	{\left( #1 \right)}		       
\newcommand  \sqparen[1]{\left[ #1 \right]}		       

\newcounter{romanum}
\renewcommand\theromanum{\textmd{\textup{(\textit{\roman{romanum}})}}}
\newenvironment{romanum}
          {\setcounter{romanum}{0}
           \begin{list}{\theromanum}
          {\usecounter{romanum}
           \setlength{\parsep}{0pt}
				\settowidth{\labelwidth}{\textmd{\textup{(\textit{viii})}}}
           \setlength{\itemsep}{2pt}}}{\end{list}}

\newcommand\s{\textsf{s}}

\newcommand\arc\smallto

\newcommand\comp{^{\textsf{c}}}

\newcommand\mP{\mathcal{P}}
\newcommand\mC{\mathcal{C}}
\newcommand\mF{\mathcal{F}}

\newcommand\mT{\mathcal{T}}

\newcommand\mI{\mathcal J}

\newcommand\flowi{(\ref{flow:i})}
\newcommand\flowii{(\ref{flow:ii})}
\newcommand\flowiii{(\ref{flow:iii})}

\newcommand\naturaliii{(\ref{naturaliii})}

\def\puttext(#1,#2)[#3]#4{\put(#1,#2){\put(-0.5,0){\makebox(0,0)[#3]{#4}}}}

\begin{document}

\title{Finding flows in the one-way measurement model}
\author{Niel \surname{de Beaudrap}}
\affiliation{IQC, University of Waterloo}
\email{jdebeaud@iqc.ca}
\thanks{supported in part by ARDA, ORDCF, MITACS, and CIAR.}

\date{\today}

\begin{abstract}
	The one-way measurement model is a framework for universal quantum computation, in which algorithms are partially described by a graph $G$ of entanglement relations on a collection of qubits.
	A sufficient condition for an algorithm to perform a unitary embedding between two Hilbert spaces is for the graph $G$, together with input/output vertices $I, O \subset V(G)$, to have a \emph{flow} in the sense introduced by Danos and Kashefi~\cite{DK05}.
	For the special case of $\abs{I} = \abs{O}$, using a graph-theoretic characterization, I show that such flows are unique when they exist.
	This leads to an efficient algorithm for finding flows, by a reduction to solved problems in graph theory.
\end{abstract}

\maketitle

\section{Introduction}

The one-way measurement model is a framework for quantum computation, first presented in~\cite{RB01,RB02}.
Transformations of quantum states in the one-way measurement model are essentially described by a sequence of single-qubit measurements (where the choice of measurement may depend on earlier measurement results in a straightforward way) performed on a many-qubit entangled state.
The many-qubit system includes some number of \emph{input qubits} $I$ in an unknown initial state, and a collection of auxiliary qubits prepared in the $\ket{+}$ state. These are operated on by a collection of entangling operations, described by a graph $G$ whose edges $uv \in E(G)$ are pairs of qubits operated on by two-qubit controlled-$Z$ operations.
After the sequence of measurements, any qubits left unmeasured still support a quantum state, and are interpreted as an \emph{output system} $O$.


Algorithms in the one-way model may be obtained by translating from the circuit model: we may decompose a unitary operation $U$ into one- and two-qubit unitaries which have known implementations in the one-way model (e.g. Hadamards, $\pi/8$ gates, and controlled-$Z$ gates), and compose the operations for these unitaries to find an algorithm for the composite unitary $U$.
We may then transform the measurement algorithm so that all of the entangling operations are performed first.

Is it possible to develop measurement algorithms without reference to the circuit model?
One proposal~\cite{BDK06} reduces the problem of implementing a unitary in the one-way model to a problem of expressing the complex coefficients of the unitary operator to be implemented in terms of sums of roots of unity, which define an entanglement graph through their ratios.
Doing this requires that one determine the order in which the measurements are to be made.
This may be done by making use of a \emph{flow} in the sense introduced in~\cite{DK05}, which is a property of just the entanglement graph and the vertex sets $I, O \subset V(G)$.

In this article, I describe how to efficiently determine whether a graph $G$ (together with input/output vertices $I$ and $O$) has a flow in the sense of~\cite{DK05}, for the special case $\abs{I} = \abs{O}$ corresponding to algorithms performing unitary transformations (as opposed to general unitary embeddings).
This is done by characterizing flows in terms of a families of vertex-disjoint paths, and proving that these are unique (when they exist) in the case of $\abs{I} = \abs{O}$.
This allows flows to be constructed efficiently, when they exist, by reduction to solved problems on directed graphs.

\section{Preliminaries}

In this section, we will review the one-way measurement model, and the concept of a flow which pertains to it.
For basic definitions in graph theory, readers may refer to Diestel's excellent text~\cite{Diestel}.

\subsubsection*{Notation and conventions.}

For a graph $G$, we write $V(G)$ for the set of vertices and $E(G)$ for the set of edges of $G$. Similarly, for a directed graph (or \emph{digraph}) $D$, we write $V(D)$ for the set of vertices and $A(D)$ for the set of directed edges (or \emph{arcs}) of $D$.
If $x$ and $y$ are adjacent, we let $xy$ denote the edge between them in a graph, and $x \arc y$ denote an arc from $x$ to $y$ in a digraph. When a graph $G$ is clear from context, we will write $x \sim y$ for the adjacency relation in $G$.
We will use the convention that digraphs may contain loops on a single vertex and multiple edges between two vertices, but that undirected graphs have neither; and $\N$ will denote the non-negative integers. 

\subsection{The one-way measurement model}

Computations in the one-way measurement model are described by a sequence of primitive operations on a set of qubits $V$.
Using the notation of~\cite{DKP04b}, the permitted operations are:
\begin{itemize}
\item
	\emph{preparation maps} $N_v$, which perpare a qubit $v \in V$ in the $\ket{+}$ state;

\item
	\emph{entangling operations} $E_{vw}$, which perform a controlled-$Z$ operation on qubits $v, w \in V$\,;

\item
	\emph{correction operations} consisting of $X$ or $Z$ operations on single qubits;

\item
	\emph{measurement operations} $M_v^\alpha$, which perform a measurement of a single qubit $v$ in an orthonormal basis of states in the equator of the Bloch sphere.
\end{itemize}
The measurements $M^\alpha$ may be described by observables of the form $\Pi_\alpha = \tfrac{1}{2} \paren{\idop - \cos(\alpha) X - \sin (\alpha) Y}$; this operator has eigenvectors $\ket{\pm_\alpha} \propto \ket{0} \pm \e^{i \alpha} \ket{1}$, with $\ket{+_\alpha}$ having eigenvalue $0$ and $\ket{-_\alpha}$ having eigenvalue $+1$.
The operator $M^\alpha_v$ represents measuring the qubit $v$ using the projector $\Pi_\alpha$, and recording the eigenvalue of the result as a bit $\s_v$ (referred to as the \emph{measurement signal})~\footnote{%
No assumption is usually made about whether the qubit $v$ exists after measurement, and so it is generally ignored as a part of any output state; however, if it does exist, complete information about its' state is provided by the values of $\alpha$ and $\s_v$.}.
Later correction or measurement operations can depend on the value of $\s_v$, which is referred to as \emph{classical feed-forward} of measurement results.

When using the measurement-based quantum computation to perform ``quantum-to-quantum'' operations (i.e. to transform quantum states), we identify two special sets of qubits: the set of \emph{input qubits} $I$, which are not operated on by a preparation map, and the set of \emph{output qubits} $O$, which are not measured.
The initial state of the qubits in $I$ may be arbitrary, and represents the input of the measurement algorithm; and the qubits of $O$ retain a final quantum state, which represents the output of the algorithm.
A valid algorithm is any sequence of the above operations with the following properties:
\begin{romanum}
\item each qubit is prepared~at most once and measured at most once;
\item no operation may depend on a measurement signal $\s_v$ before the qubit $v$ has been measured;
\item the first operation performed on a qubit $v \in I\comp = V \setminus I$ is a preparation map;
\item the last operation performed on a qubit $v \in O\comp = V \setminus O$ is a measurement.
\end{romanum}

Algorithms can be condensed by allowing the measurement operations to depend on the results of previous measurements, and by performing all entanglement operations towards the beginning of the algorithm.
Let $s$ be a boolean expression: then, using the equalities
\begin{subequations}
\label{eqn:correctionCommute}
\begin{align}
		E_{vw} X_v^s 
	\;=&\;
		X_v^s Z_w^s E_{vw}	\,,
 	\\
		M^\alpha_v X_v^s
	\;=&\;
		M^{\alpha \cdot (- 1)^{\scriptstyle s}}_v	,
 	\\
		M^\alpha_v Z_v^s
	\;=&\;
		M^{\alpha + \pi \cdot s}_v	,
\end{align}
\end{subequations}
we may postpone all correction operations until the end of the algorithm, and perform all preparation/entangling operations at the beginning.
This allows us to describe measurement algorithms in the usual way with the preparation of an entangled resource (an \emph{open graph state}, depending on the initial state of the input qubits $I$), with measurement and correction operations performed on it;
and where the angles of measurements may depend on the signals produced by earlier measurements.
Note that measurements of the form $M^{\alpha + \pi \cdot s}_v$ are performed with respect to the same basis as $M^\alpha$, but with the two basis elements $\ket{\pm_\alpha}$ interchanged whenever $s \equiv 1 \pmod{2}$.
Rather than changing the angle of measurement by $\pi$ depending on the value of $s$, we may add the value of the expression $s$ to the measurement result $\s_v$ to obtain a modified result $\s'_v$.
Equivalently, in the algebraic representation of the measurement algorithm, we may substitute $\s_v$ wherever it occurs in a correction of measurement dependency with the expression $\s_v + s$: this substitution is called \emph{signal shifting} in~\cite{DKP04b}.


Thus, without loss of generality, we may describe unitary transformations using measurement-based algorithms where, if all measurement results are $0$, no adaptation of the measurement angles is required, nor are corrections on the final output system. This prompts the question of whether we can design algorithms \emph{in terms of} the behaviour when all the measurement results are $0$.

\subsection{Flows in the one-way measurement model}
\label{sec:flowsMeasModel}

Considering the outward differences between the one-way measurement model and the circuit model could lead to new techniques for developing quantum algorithms, as proposed by~\cite{BDK06}.
However, an apparent obstacle to directly devising algorithms in the one-way model is that measurements bases and final corrections on the output qubits may depend on many measurement signals.
These details are essential, and raises the questions of the order in which measurements are to be performed, and what measurement dependencies are required.
Unfortunately, this complicates any direct understanding of how to perform operations in the one-way measurement model, without e.g.~translating from the circuit model.

Given a graph, and a collection of measurements which yield a particular operation in the special case where all measurement results come out $0$, the problem described above can be solved if we can determine a measurement sequence and signal dependencies from the entanglement graph itself, along with the sets of input and output vertices.
\label{discn:inferByproduct}
In such a sequence of operations, we may treat each measurement $M^\alpha_v$ as post-selecting the state of the whole system so that $v$ is in the state $\ket{+_\alpha}$; should the opposite result occur, the final corrections and the intermediate changes in measurement angles are equivalent to performing a correction immediately after the measurement on $v$ to bring about the state that would have arisen had $\ket{+_\alpha}$ been the result.
We may do this if we can infer suitable by-product operations for each measurement: this can be done if the entanglement graph has a flow property introduced in~\cite{DK05}.

\begin{definition}
	\label{dfn:causalFlow}
	A \emph{geometry} $(G, I, O)$ is an entanglement graph $G$, together with subsets $I, O \subset V(G)$ representing the sets of input and output vertices of a measurement algorithm.
	A \emph{flow} on $(G,I,O)$ is an ordered pair $(f, \preceq)$, with a function $f: O\comp \to I\comp$ and a partial order~\footnote{A partial order is a binary relation which is reflexive ($x \preceq x$ for all $x$), transitive ($x \preceq y$ and $y \preceq z$ imply $x \preceq z$), and anti-symmetric ($x \preceq y$ and $y \preceq x$ only if $x = y$).} $\preceq$ on $V(G)$, such that the conditions
	\begin{subequations}
	\begin{align}
		\label{flow:i}
			x &\sim f(x)
		\\
		\label{flow:ii}
			x &\preceq f(x)
		\\
		\label{flow:iii}
			y \sim f(x) \;&\implies\; x \preceq y
	\end{align}
	\end{subequations}
	hold for all vertices $x \in O\comp$ and $y \in V(G)$. We will refer to $f$ as the \emph{successor function} of the flow, and $\preceq$ as the \emph{causal order} of the flow.
\end{definition}

\begin{figure}[t]
	\begin{center}
			\includegraphics{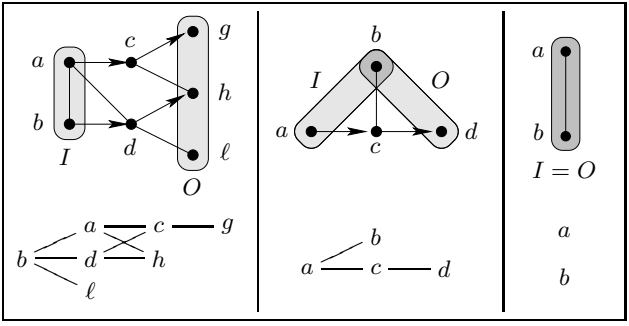}
	\end{center}
	\caption{\label{fig:examplesFlow}
		Examples of geometries with flows.
		Arrows indicate the action of a successor function $f: O\comp \to I\comp$,\/ along \emph{undirected} edges.
		Causal orders $\preceq$ for each example are given by Hasse diagrams (read from left to right).
		In the right-most example, the two vertices $a$ and $b$ are \emph{incomparable}, i.e. there is no order relation between them.}
\end{figure}

Figs.~\ref{fig:examplesFlow} and~\ref{fig:exampleNoFlow} illustrate examples of geometries with and without flows.
The conditions \flowi~--~\flowiii\ are meant to capture a simple set of conditions, independent of the angles of the measurements, which are sufficient to determine how to adapt measurement angles and perform corrections to perform unitary transformations.
Specifically, it captures when the by-product operations for each measurement $M^\alpha_x$ can be considered to consist of a single $X$ operation on some qubit $f(x)$, and $Z$ operations on each qubit $y \sim f(x)$.

The above is not a \emph{necessary} condition for a geometry $(G,I,O)$ to permit unitary evolution independently of the measurements performed (see~\cite{BKMP07} for a generalization), but it is sufficient.
The partial order $x \preceq y$ then represents when such a byproduct operation $C_x$ for the measurement on $x$ acts non-trivially on $y$.
If we insert the operation $C_x$ after the measurement on $x$ (deleting the Pauli operation on $x$ itself) in a measurement algorithm consisting only of preparation maps, entanglers, and measurements, these corrections will be absorbed into the measurements on the qubit $f(x)$ and each of the qubits $y \sim f(x)$.
Performing such substitutions/absorptions for all of the qubits to be measured, the order $\preceq$ then describes chains of measurement dependencies~\footnote{%
For a qubit $x \in O\comp$ and a qubit $y \sim f(x)$, the by-product operation for the measurement on $x$ does not change the basis of measurement for $y$, but it does change the significance of a $\ket{+_\alpha}$ or $\ket{-_\alpha}$ at $y$ for future measurements or corrections, as a $Z$ correction on $y$ interchanges these two states.
So, more precisely, we have $x \preceq y$ if the either measurement angle for $f(y)$ may change sign depending on the signal $\s_x$, or if $y \in O$ and there is an $X$ or $Z$ correction on $y$ which depends on $x$.}.

Thus, having a flow allows us to infer a sequence of measurements, and suitable dependencies for those measurements, providing a solution to the problem described towards the beginning of this section.
This makes it easier to design quantum algorithms directly in the one-way measurement model, by obtaining complete sequencess of measurement operations from only partial information.
This motivates the problem of efficiently determining when a geometry has a flow.

\begin{figure}[t]
	\begin{center}
			\includegraphics{fig2.epsi}
	\end{center}
	\caption{\label{fig:exampleNoFlow}
		A geometry with no flow.
		Also shown is a particular injection $f: O\comp \to I\comp$, and the coarsest pre-order satisfying conditions \flowii\ and \flowiii\ --- see the discussion on page~\pageref{discn:cycleGeometry} preceding Definition~\ref{dfn:viciousCircuit}.}
\vspace{1.7em}
\end{figure}

\section{Characterizing flows in graph-theoretic terms}

In order to determine whether a geometry $(G,I,O)$ has a flow, it is useful to understand the sorts of structures which are induced or forbidden in $G$ by the presence of a flow.
We begin with a restriction of the concept of a \emph{path cover} to geometries:

\begin{definition}
	\label{dfn:pathCover}
	Let $(G,I,O)$ be a geometry. A collection $\mC$ of (possibly trivial) directed paths in $G$ is a \emph{path cover} of $(G,I,O)$ if
	\begin{romanum}
	\item
		each $v \in V(G)$ is contained in exactly one path (i.e. the paths cover $G$ and are vertex-disjoint);
	\item
		each path in $\mC$ is either disjoint from $I$, or intersects $I$ only at its initial point;
	\item
		each path in $\mC$ intersects $O$ only at its final point.
	\end{romanum}
\end{definition}

In the case $\abs{I} = \abs{O}$, a path cover of $(G,I,O)$ will just be a collection of vertex-disjoint paths from $I$ to $O$ which covers all the vertices of $G$.

For a flow $(f,\preceq)$, there is a natural connection between the successor function $f$ and path covers for the geometry $(G,I,O)$, which we capture in the following Lemma:

\begin{lemma}
	\label{lemma:orbitPathCover}
	Let $(f, \preceq)$ be a flow on a geometry $(G,I,O)$. Then there is a path cover $\mP_f$ of $(G,I,O)$, where $x \arc y$ is an arc in some path of $\mP_f$ if and only if $y = f(x)$.
\end{lemma}
\begin{proof}
	Let $(f,\preceq)$ be a flow on $(G,I,O)$.
	Suppose that $f(x) = f(y)$ for some $x, y \in O\comp$.
	By condition~\flowi, we have $y \sim f(y) = f(x)$; and by condition~\flowiii, we have $x \preceq y$.
	Similarly, we have $y \preceq x$, so $x = y$. Thus $f$ is an injective function.

	Define a digraph $P$ on the vertices of $G$, and with arcs $x \arc f(x)$ for $x \in O\comp$.
	Because $f$ is both a function and injective, every vertex in $P$ has maximal
	out-degree and maximal in-degree $1$. Thus, $P$ is a collection of vertex-disjoint dipaths,
	dicycles, and closed walks of length $2$. As well, for every arc $(x \arc y) \in A(P)$,
	we have $x \preceq y$; by induction, $x \preceq z$ whenever there is a dipath from $x$ to
	$z$ in $P$. Then if $x$ and $z$ are such that there are dipaths from $x$ to $z$
	and from $z$ to $x$, then $x \preceq z$ and $z \preceq x$, in which case
	$x = z$ and the dipaths are trivial. Thus, $P$ is acyclic, so $P$ consists
	entirely of vertex-disjoint dipaths.

	Let $\mP_f$ be the collection of maximal dipaths in $P$. We show that $\mP_f$ satisfies
	each of the criteria of Definition~\ref{dfn:pathCover}:
	\begin{romanum}
	\item
		Any vertex $v$ which is neither in $\dom(f)$ nor $\img(f)$ will be isolated in $P$:
		then, the trivial path on $v$ is an element of $\mP_f$. All other vertices are in
		either $\dom(f)$ or $\img(f)$, and so are contained in a non-trivial path of
		$\mP_f$. As these paths are vertex-disjoint, each vertex is contained in
		exactly one path.

	\item
		Each vertex in $I$ has in-degree $0$, and so may only occur at the beginning of any
		path in $\mP_f$.

	\item
		The vertices in $P$ which have out-degree $0$ are precisely the output vertices $O$: therefore
		one must occur at the end of every path, and they may only occur at the end of paths in $\mP_f$.
	\end{romanum}
	Then $\mP_f$ is a path cover, whose paths contain only arcs $x \arc f(x)$, as required.
\end{proof}

It will prove useful to discuss functions $f$ which are not necessarily the successor function of a flow $(f, \preceq)$, but which nonetheless are related to a path cover in the sense of Lemma~\ref{lemma:orbitPathCover}.
Thus, we will extend our usage of the term \emph{successor function} to include the following definition:

\begin{definition}
	\label{dfn:successorFn}
	Let $\mC$ be a path cover for a geometry $(G,I,O)$. Then the \emph{successor function
	of $\mC$} is the unique $f: O\comp \to I\comp$ such that $y = f(x)$ if and only
	if $x \arc y$ is an arc in some path of $\mC$. If a function $f: O\comp \to I\comp$ is a successor
	function of \emph{some} path-cover of $(G,I,O)$, we may call $f$ a \emph{successor
	function of $(G,I,O)$}.
\end{definition}
In the case where $\abs{I} = \abs{O}$, the successor function of a geometry $(G,I,O)$ is bijective.
This allows us to define the additional useful terminology:
\begin{definition}
	Let $\mC$ be a path cover for a geometry $(G,I,O)$ with $\abs{I} = \abs{O}$. The
	\emph{predecessor function} of $\mC$ is the unique $g: I\comp \to O\comp$ such that
	$g(y) = x$ if and only if $x \arc y$ is an arc in some path of $\mC$.
\end{definition}

Given that the successor function of a flow for $(G,I,O)$ induces a path cover,
one might think of also trying to obtain a flow from a path cover. There is an
obvious choice of binary relation which we would like to consider, which satisfies
conditions~\flowii\ and~\flowiii:
\begin{definition}
	\label{dfn:naturalPreorder}
	Let $f$ be a successor function for $(G,I,O)$. The \emph{natural pre-order~\textup{\footnote{%
	A pre-order is a binary relation which is reflexive and transitive, but not necessarily antisymmetric.}}
	$\preceq$ for $f$} is the transitive closure on $V(G)$ of the conditions
	\begin{subequations}
	\begin{align}
			\label{naturali}
			x &\preceq x
		\\
			\label{naturalii}
			x &\preceq f(x)
		\\
			\label{naturaliii}
			y \sim f(x) \;\;&\implies\;\; x \preceq y
	\end{align}
	\end{subequations}
	for all $x, y \in V(G)$.
\end{definition}
Recall from Section~\ref{sec:flowsMeasModel} the description of causal orders $\preceq$ for flows in terms of chains of measurement dependencies arising from byproduct operations: we have $x \preceq y$ if there is a sequence of vertices $x = z_0, z_1, \cdots, z_\ell = y$ such that the byproduct operator for the measurement on $z_j$ acts non-trivially on $z_{j+1}$ for each $0 \le j < \ell$.
We are then interested in when a natural pre-order $\preceq$ is antisymmetric, in which case it provides a well-defined order for measurements.

It is easy to show that the natural pre-order $\preceq$ for $f$ is a partial order if and
only if $f$ is the successor function of a flow. If $\preceq$ is a partial order, it
will be the coarsest partial order such that $(f, \preceq)$ is a flow. However, it is
easy to construct geometries and successor functions $f$ for which the natural pre-order $\preceq$
is not a partial order. One example \label{discn:cycleGeometry} is the geometry $(G,I,O)$ illustrated
in Fig.~\ref{fig:exampleNoFlow} on page~\pageref{fig:exampleNoFlow}, with $G$ equal to the cycle $C_6 = a_0 b_0 a_1 b_1 a_2 b_2
a_0$, $I = \ens{a_0, a_1, a_2}$, and $O = \ens{b_0, b_1, b_2}$. For any successor 
function $f$ on this geometry, condition~\naturaliii\ forces either $a_0 \preceq a_1 \preceq a_2 \preceq
a_0$ or $a_0 \succeq a_1 \succeq a_2 \succeq a_0$ to hold. Because $a_0$, $a_1$, and
$a_2$ are distinct, such a relation $\preceq$ is not antisymmetric, so it is not a partial order.
In this case, we have not only a cyclic graph, but a cycle of relationships induced
by condition~\naturaliii. The following definitions are aimed to characterize these cyclic
relations in terms of closed walks.
\begin{definition}
	\label{dfn:viciousCircuit}
	Let $(G,I,O)$ be a geometry, and $\mF$ a family of directed paths in $G$. A walk
	$W = u_0 u_1 \cdots u_\ell$ is an \emph{influencing walk}~\footnote{Influencing walks are closely
	related to \emph{walks which alternate with respect to $\mF$} as defined in Section~3.3
	of~\cite{Diestel}. The term ``influencing walk'' is due to Broadbent and Kashefi, who examine
	their role in the depth complexity of unitaries in the one-way measurement model~\cite{BK07},
	and identified them as objects of interest after reading an earlier draft of this article.}
	for $\mF$ if it is a concatenation of zero or more paths (called \emph{segments}
	of the influencing walk) of the following two types:
	\begin{itemize}
	\item
		$x \arc y$, where this is an arc in some path of $\mF$;
	\item
		$x \arc z \arc y$, where $x \arc z$ is an arc in some path of $\mF$ and $y z \in E(G)$ is not covered by $\mF$.
	\end{itemize}
	A \emph{vicious circuit} for $\mF$ is a closed influencing walk for $\mF$ with at least
	one segment.
\end{definition}
A non-trivial influencing walk $W$ of $\mF$ must start with an arc in some path of $\mF$; and that of any two consecutive edges of $W$, at least one is an arc in some path of $\mF$.
Then, it is easy to see that the decomposition of $W$ into its' segments is unique: the initial segment is of the first type if and only if the first two edges are arcs of $\mF$, is of the second type otherwise.
The entire walk can be decomposed recursively in this fashion.

\begin{definition}
	\label{dfn:causalPathCover}
	Let $(G,I,O)$ be a geometry. A path cover $\mC$ for $(G,I,O)$ is a \emph{causal path cover} if $\mC$ does not have any vicious circuits in $G$.
\end{definition}

The two types of segments which build an influencing walk correspond to the flow conditions~\flowii\ and~\flowiii:
influencing walks again represent chains of dependencies induced by byproduct operations. Specifically:

\begin{lemma}
	\label{lemma:influencingNatlPreorder}
	Let $\mC$ be a path cover for $(G,I,O)$ with successor function $f$, and let $\preceq$ be the natural pre-order of $f$.
	Then $x \preceq y$ if and only if there is an influencing walk for $\mC$ from $x$ to $y$.
\end{lemma}
\begin{proof}
	To show that $x \preceq y$ if there is an influencing walk from $x$ to $y$, we proceed
	by induction on the number of segments of the influencing walk. If the number of
	segments of the influencing walk is zero, then $x = y$, in which case $x \preceq y$.
	Otherwise, suppose the proposition holds for all influencing walks for $\mC$ with fewer
	than $n$ segments for some $n \in \N$, and that there is an influencing walk $W = x
	u_1 \cdots u_\ell y$ from $x$ to $y$ (for some vertex-sequence $\paren{u_j}_{j =
	1}^\ell$) which has $n$ segments.
	\begin{itemize}
	\item
		If the final segment of $W$ is $u_\ell y$, then $x u_1 \cdots u_\ell$ is an
		influencing 	walk of $n-1$ segments, so $x \preceq u_\ell$. Because we also
		have $y = f(u_\ell)$, from the definition of the natural pre-order we have $x
		\preceq y$.
		
	\item
		If the final segment of $W$ is $u_{\ell-1} u_{\ell} y$, then $x u_1 \cdots
		u_{\ell-1}$ is an influencing walk of $n-1$ segments, so $x \preceq
		u_{\ell-1}$. Because we also have $y \sim u_\ell = f(u_{\ell-1})$, from
		the definition of the natural pre-order we have $x \preceq y$.
	\end{itemize}

	Conversely: from the definition of $\preceq$ as a transitive closure, if $x \preceq y$
	for some $x,y \in V(G)$, there is a sequence of vertices $\paren{u_j}_{j = 0}^\ell$
	for some $\ell \in \N$ such that $u_0 = x$, $u_m = y$, and either $u_{j+1} = f(u_j)$
	or $u_{j+1} \sim f(u_j)$ holds for each $0 \le j < \ell$. Then, define the paths
	\begin{align}
			\sigma_j
		\;\;=&\;\;
			\begin{cases}
				\;	u_j u_{j+1}	\,,			&	\text{if $u_{j+1} = f(u_j)$}	\\
				\;	u_j f(u_j) u_{j+1}\,,	&	\text{if $u_{j+1} \sim f(u_j)$}
			\end{cases}\quad
	\end{align}
	for each $0 \le j < \ell$: the walk $\sigma_0 \sigma_1 \cdots \sigma_\ell$ obtained from
	concatenating these paths is a walk from $x$ to $y$, and in particular an influencing
	walk.
\end{proof}

This equivalence allows us to characterize flows in terms of paths and circuits in the graph:

\begin{theorem}
	\label{thm:graphth-charn}
	Let $(G,I,O)$ be a geometry with path cover $\mC$, $f$ be the successor function of
	$\mC$, and $\preceq$ be the natural pre-order for $f$\,. Then the following are
	equivalent:
	\begin{romanum}
	\item
		\label{graph-charn-i}
		$\mC$ has no vicious circuits;
	\item
		\label{graph-charn-ii}
		$\preceq$ is a partial order;
	\item
		\label{graph-charn-iii}
		$(f,\preceq)$ is a flow.
	\end{romanum}
	In particular, a geometry has a flow iff it has a causal path cover.	
\end{theorem}
\begin{proof}
	By construction, $(f,\preceq)$ fails to be a flow if and only
	if $\preceq$ is not a partial order (i.e. if and only if it is not anti-symmetric). Thus
	$\ref{graph-charn-ii} \iff \ref{graph-charn-iii}$.

	If $\preceq$ is not anti-symmetric, then there are distinct $x, y \in V(G)$ such
	that $x \preceq y$ and $y \preceq x$: by Lemma~\ref{lemma:influencingNatlPreorder},
	there is then an influencing walk $W$ with at least one segment from $x$ to $y$,
	and also an influencing	walk $W'$ with at least one segment from $y$ to $x$. Then
	$W W'$ is an influencing walk with at least two segments from $x$ to itself, and
	is therefore a vicious circuit for $\mC$; then $\mC$ is not a causal path cover.
	Conversely, if $\mC$ is not a causal path cover, then there is a vicious circuit
	$x u_1 u_2 \cdots u_{\ell-1} x$ for $\mC$: if $u_2 = f(u_1)$, then
	$x u_1$ and $u_1 u_2 \cdots x$ are both influencing walks, in which case
	$x \preceq u_1 \preceq x$; otherwise, $x u_1 u_2$ and $u_2 \cdots u_{\ell-1} x$
	are both influencing walks for $\mC$, in which case $x \preceq u_2 \preceq x$.
	Thus $\ref{graph-charn-i} \iff \ref{graph-charn-ii}$.
\end{proof}

Characterizing flows in terms of causal path covers allows us to shift the
emphasis from the constructibility of a causal order $\preceq$ to the absence of vicious
circuits. By using successor and predecessor functions, we may show that requiring
vicious circuits to be absent for a path cover yields a strong uniqueness result:

\begin{theorem}
	\label{thm:uniqueFamilyPaths}
	Let $(G,I,O)$ be a geometry such that $\abs{I} = \abs{O}$. If $(G,I,O)$ has a causal path cover $\mC$, then $\mC$ is also the only
	maximum collection of vertex-disjoint dipaths from $I$ to $O$.
\end{theorem}
\begin{proof}
	Suppose that $\mC$ is a path cover for $(G,I,O)$ with successor function $f: O\comp
	\to I\comp$, and suppose there is a maximum-size collection $\mF$ of vertex-disjoint
	$I$ -- $O$ dipaths which differs from $\mC$. Let $S \subset V(G)$ be the set of
	vertices not covered by $\mF$: because $\abs{\mF} = \abs{\mC} = \abs{I} = \abs{O}$,
	we have $S \inter I = \vide$ and $S \inter O = \vide$, in which case $\mF$ is a path
	cover for the geometry $(G \setminus S, I, O)$. Then, let $g': (I\comp \setminus S)
	\to (O\comp \setminus S)$ be the predecessor function of $\mF$ as a path cover of
	$(G \setminus S, I, O)$. 
	
	Because $\mC$ and $\mF$ differ, there must exist a vertex $x \in O\comp$ such that
	$x \arc f(x)$ is not an arc in some path of $\mF$. Note also that for $v \in \dom(f)$,
	$f(v) \notin \dom(f)$ holds only if $f(v) \in O \setminus I \subset \dom(g')$; that
	is, $f(v) \in \dom(f) \union \dom(g')$. Then, define a vertex sequence
	$\paren{u_j}_{j \in \N}$ in $G$ by setting $u_0 = x$, $u_1 = f(x)$, and 
	\begin{align}
			\label{eqn:inflWalkRecurrance}
			u_{j+1}
		\;=&\;
			\begin{cases}
				\; f(u_j)	\;,	&\!\!	\text{$u_j \in S$ or $u_j \ne f(u_{j-1})$}	\\[.5ex]
				\; g'(u_j)	\;,	&\!\!	\text{$u_j \notin S$ and $u_j = f(u_{j-1})$}
			\end{cases}
	\end{align}
	for all $j \ge 1$. Fig.~\ref{fig:exampleInflConstr} illustrates this construction.

\begin{figure}[t]
	\begin{center}
		\includegraphics{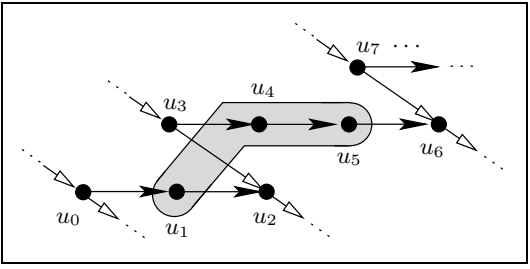}
	\end{center}
	\caption{\label{fig:exampleInflConstr}
 		An influencing walk for a path cover $\mC$ (solid arrows) induced by another maximum collection $\mF$ of vertex-disjoint paths from $I$ to $O$ (hollow arrows).
		The shaded area is a subset of the set $S \subset V(G)$ not covered by $\mF$.}
\end{figure}

	Clearly $u_j \sim u_{j+1}$ for all $j \in \N$. We also have $u_0 \arc u_1$ an
	arc in some path of $\mC$, and for any $j \ge 1$ such that $u_j \arc u_{j+1}$ is not an arc
	of $\mC$, it follows that $u_j \ne f(u_{j-1})$, in which case we have
	$u_{j+1} = f(u_j)$, which implies $u_j \arc u_{j+1}$ is an arc in some path of $\mC$.
	Then for any $N \in \N$, the walk $W_N = u_0 u_1 \cdots u_N$ is an influencing walk
	in $G$.
	
	Because $G$ is a finite graph, the Pigeon Hole Principle implies that there must be integers
	$m, m' \in \N$ with $m < m'$, $u_m = u_{m'}$, and $u_{m-1} = u_{m'-1}$.
	Because $W_{m'}$ is an influencing walk, at least one of $u_{m-1} \arc u_m$ or
	$u_m \arc u_{m+1}$ is an arc in some path of $\mC$. In the former case, the closed walk
	$u_{m-1} u_m \cdots u_{m'-1}$ is an influencing walk, and thus a vicious
	circuit; otherwise, $u_m u_{m+1} \cdots u_{m'}$ is an influencing walk,
	and thus a vicious circuit. In either case, there exists a vicious circuit for
	$\mC$, in which case $\mC$ is not a causal path cover.

	Thus, if $\mC$ is a causal path cover, there can be no such vertex
	sequence $\paren{u_j}_{j \in \N}$ as defined above, and so there can be no maximum
	family of vertex-disjoint $I$ -- $O$ paths $\mF$ which differs from $\mC$.
\end{proof}

Note that for $\abs{I} = \abs{O}$, because a causal path cover of $(G,I,O)$ is unique if it exists,
and the successor function of any flow will also be the successor function of a causal path cover,
there is at most one successor function $f$ which yields a flow for $(G,I,O)$. Because the natural pre-order
$\preceq$ for $f$ is coarser than any other valid causal order for $f$, it too is unique. Then, a
``mimumum-depth'' flow for a geometry $(G,I,O)$ is unique in the case $\abs{I} = \abs{O}$.

\section{Finding flows efficiently when \protect{$\abs{I} = \abs{O}$}}
\label{sec:polyTimeAlg}

Theorem~\ref{thm:uniqueFamilyPaths} allows us to reduce the problem of finding a flow for $(G,I,O)$ when $\abs{I} = \abs{O}$ to finding a maximal collection $\mC$ of vertex-disjoint $I$ -- $O$ paths, and then determining whether or not $\mC$ has vicious cycles.
Both steps can be expressed in terms of solved problems in directed graphs, and both can be solved in time $O(km)$, where $k = \abs{I} = \abs{O}$, and $m = \abs{E(G)}$. An upper bound on the number of edges that a geometry may have if it has a flow~\cite{BP07} allows us to further bound this by $O(k^2 n)$, where $n = \abs{V(G)}$.
In this section, I give an outline for the solution of these results to show that a flow can be found efficiently when $\abs{I} = \abs{O}$.~\footnote{For a detailed description of algorithms to efficiently find flows, readers should refer to~\cite{B06b}, which presents complete algorithms and proofs of their correctness without assuming any prior knowledge of graph-theoretic algorithms.} 

\subsection{Finding a path cover for $(G,I,O)$}

Finding a path cover for $(G,I,O)$ can be reduced to an instance of \emph{network flows}. A \emph{network} is
a directed graph $N$ with a designated \emph{source vertex} $r$ and \emph{sink vertex} $s$, and a capacity
function $c: A(N) \to \N$ representing the maximum rate at which some substance can pass through each arc.
An \emph{integral $r$ -- $s$ network flow} is a function $\phi: A(N) \to \N$ such that $\phi(a) \le c(a)$ for all
$a \in A(N)$, and where the ``net flow'' into a vertex $x \in V(N)$,
\begin{align}
	\Phi(x)
	\;=&\;	
		\sqparen{\sum_{\substack{\scriptscriptstyle (u \arc x) \\\scriptscriptstyle  \in A(N)}} \phi(u \arc x)}
	-
		\sqparen{\sum_{\substack{\scriptscriptstyle (x \arc v) \\\scriptscriptstyle  \in A(N)}} \phi(x \arc v)}
\end{align}
is zero for $x \notin \ens{r,s}$. The \emph{value} of the network flow $\phi$  is $\Phi(r)$.

We may start the reduction to network flows by augmenting the entanglement graph $G$ to a graph $G'$, adding
a vertex $r$ which is adjacent to every vertex of $I$, and a vertex $s$ which is adjacent to every vertex of
$O$. Any collection of vertex-disjoint $I$--$O$ paths then corresponds naturally to a collection of ``internally
disjoint'' paths from $r$ to $s$ of the same size. By a construction presented in Section~8.3 of~\cite{CH91},
we can then efficiently construct from $G'$ a network $N$ with source $r$ and sink $s$, such that
every integral $r$ -- $s$ network flow $\phi$ can be used to construct a collection of $\Phi(r)$ internally disjoint
paths from $r$ to $s$ in $G'$. It then suffices to find a maximum integral network flow for $N$.
This is a well-studied problem: in the case where all edges have capacity $1$,
the Ford-Fulkerson algorithm (see e.g.~\cite{CLRS}, Section~26.2) runs in time $O(k' m')$, where
$k' \le \abs{I} = \abs{O}$ is the value of the maximum network flow, and where
$m' = O(\abs{E(G)})$ is the number of arcs in the network $N$.

Having found a maximum-size collection $\mF$ of vertex-disjoint paths, we may determine if
$\mF$ is a path cover simply by verifying that it covers all vertices: this may be done
in time $O(\abs{V(G)})$. If $\mF$ is not a path cover, $(G,I,O)$ has no flow by
Theorems~\ref{thm:graphth-charn} and~\ref{thm:uniqueFamilyPaths}.

\subsection{Determining a causal order}

To determine whether or not a path cover $\mC$ (with successor function $f$) for
$(G,I,O)$ has vicious circuits, we may create the digraph $\mI_f$ whose vertices are
those of $G$, and where $(x \arc y) \in A(\mI_f)$ iff there is an influencing walk
for $\mC$ of at most one segment from $x$ to $y$. Then, $\mC$ has vicious circuits
iff $\mI_f$ contains a directed cycle. Tarjan's algorithm (see e.g.~\cite{N95}, Section 3.1)
is a simple algorithm for determining the \emph{strongly connected components} of a directed
graph $D$: the equivalence classes of vertices which are mutually reachable by directed walks.
In any circuit of $\mI_f$, all of the vertices are mutually reachable; then, we can use
Tarjan's algorithm on $\mI_f$ to determine whether $\mC$ is a causal path cover.
If $\mI_f$ contains two mutually reachable vertices, $(G,I,O)$ has no flows
by Theorems~\ref{thm:graphth-charn} and~\ref{thm:uniqueFamilyPaths}.

Because the natural pre-order $\preceq$ for $f$ is characterized by influencing
walks for $\mC$, we have $x \preceq y$ iff there is a directed path from $x$
to $y$ in $\mI_f$. Then, the problem of computing $\preceq$ is equivalent to
the problem of computing the \emph{transitive closure} of $\mI_f$: the directed
graph $\mT_f$ in which there is an arc from $x$ to $y$ iff there is a non-trivial
directed walk in $\mI_f$ from $x$ to $y$. The transitive closure can
also be computed by a modification of Tarjan's algorithm: then, $\preceq$
can computed at the same time as we are determining whether $\mI_f$ contains
directed cycles (i.e. whether or not $\preceq$ is anti-symmetric). 

Each path of $\mC$ is totally ordered by the pre-order $\preceq$: then, we can represent
the relation $\preceq$ efficiently through a \emph{chain decomposition} --- for each $x \in V(G)$,
we store the minimal element $y_P$ in each path $P \in \mC$ such that $x \preceq y_P$. From
Theorem~3.11 of~\cite{N95}, we can compute $\preceq$ in time $O(km)$, where $k = \abs{\mC} =
\abs{I} = \abs{O}$ and $m = \abs{E(G)}$. As remarked in the previous paragraph, we may determine
whether $\preceq$ is a partial order at the same time: if it is, $(f,\preceq)$ is a flow
for $(G,I,O)$; otherwise, $(G,I,O)$ has no flows.

\subsection{Eliminating geometries with too many edges}

For $n = \abs{V(G)}$ and $k = \abs{O}$ (but without requiring that $I$ and $O$ have the same number of elements), an extremal result~\cite{BP07} shows that any geometry $(G,I,O)$ which has a flow has at most $kn - \binom{k+1}{2}$ edges.
Thus, if $G$ has more than this number of edges, it cannot have a flow.
We can use this as a preliminary test for any geometry when deciding if it has a flow, aborting if $G$ has too many edges: if a geometry passes this test, the subroutines for finding the successor function $f$ and the partial order $\preceq$ described above run in time $O(k^2 n)$.

\section{Summary and Open Questions}

Flows are a tool for analyzing the underlying geometry of measurement algorithms,
which may make it feasible to develop algorithms in the one-way model without direct reference
to the circuit model. We have seen how they can be characterized and efficiently found using
tools of graph theory, in the special case where the input and output systems have the same
number of qubits.

One direction in which this work could be generalized is by considering the generalization of flows (called ``gflows'') presented by Browne, Kashefi, Mhalla, and Perdrix~\cite{BKMP07} describe a which accommodate more complex byproduct operations for measurements.
An efficient algorithm for finding such gflows would be a substantial advance in the line of investigation of finding appropriate byproduct operations.

Another question is whether similar work can be done in ``classical-to-classical'' measurement-based quantum computing, wherein all qubits are prepared and measured.
Is it possible to find sequences for measurement, and appropriate measurement dependencies, so that complete algorithms for universal quantum computation (including the final measurements, and without initial states other than $\ket{+}^{\!\ox n}$) can be obtained in the one-way measurement model only from partial information (the entanglement graph and the observables to be measured)?

\section{Acknowledgements.}

I would like to thank Elham Kashefi, who interested me in the problem of efficiently
finding flows for geometries and for helpful discussions; Donny Cheung and
Anne Broadbent, who provided useful insights; and Rob Raussendorf and Lana Sheridan,
for their feedback on the presentation of the results of this paper.

\bibliographystyle{amsalpha}

\end{document}